\documentclass[review, 12pt]{elsarticle}

\usepackage{lineno}

\usepackage{hyperref}
\hypersetup{colorlinks  = true}

\journal{Accepted in \textbf{Respiratory Physiology \& Neurobiology 294:103769}}
\biboptions{authoryear}
\usepackage{amsmath}
\usepackage{caption}
\usepackage{subcaption}
\usepackage{graphicx}
\usepackage{upgreek}
\usepackage{geometry}
\usepackage{array}

\usepackage{setspace}
 
\usepackage{upgreek} 

\usepackage{hyperref}

\usepackage[T1]{fontenc}
\usepackage{kpfonts}

\usepackage{etoolbox}
\makeatletter
\patchcmd{\ps@pprintTitle}
  {Preprint submitted to}
  {}
  {}{}
\makeatother

\begin{document}

\begin{frontmatter}
    
	\title{Inhalation and deposition of spherical and pollen particles after middle turbinate resection in a human nasal cavity\tnoteref{t1}}
    \tnotetext[t1]{\url{http://doi.org/10.1016/j.resp.2021.103769}}

	\author[label1]{Kiao Inthavong \corref{cor1}}
	\author[label1]{Yidan Shang}
	\author[label2]{John M. DelGaudio}
	\author[label2]{Sarah K Wise}
	\author[label2]{Thomas S Edwards}
	\author[label3]{Kimberly Bradshaw}
	\author[label3]{Eugene Wong}
	\author[label3]{Murray Smith}
	\author[label3,label4]{Narinder Singh}
	\cortext[cor1]{Corresponding Author: kiao.inthavong@rmit.edu.au}
	\address[label1]{Mechanical \& Automotive Engineering, School of Engineering, RMIT University, Bundoora, Victoria 3083, Australia}
	\address[label2]{Department of Otolaryngology-Head and Neck Surgery
		Emory University, Atlanta, Georgia USA}
	\address[label3]{Faculty of Medicine \& Health, The University of Sydney, NSW 2006, Australia}
	\address[label4]{Department of Otolaryngology, Head and Neck Surgery, Westmead Hospital, Westmead, NSW 2145, Australia}

{\setstretch{1.0}	
	\begin{abstract}
Middle turbinate resection significantly alters the anatomy and redistributes the inhaled air. The superior half of the main nasal cavity is opened up, increasing accessibility to the region. This is expected to increase inhalation dosimetry to the region during exposure to airborne particles. This study investigated the influence of middle turbinate resection on the deposition of inhaled pollutants that cover spherical and non-spherical particles (e.g. pollen). A computational model of the nasal cavity from CT scans, and its corresponding post-operative model with virtual surgery performed was created. Two constant flow rates of 5L/min, and 15L/min were simulated under a laminar flow field. Inhaled particles including pollen (non-spherical), and a spherical particle with reference density of 1000kg/m$^3$ were introduced in the surrounding atmosphere. The effect of surgery was most prominent in the less patent cavity side, since the change in anatomy was proportionally greater relative to the original airway space. The left cavity produced an increase in particle deposition at a flow rate of 15L/min. The main particle deposition mechanisms were inertial impaction, and to a lesser degree gravitational sedimentation. The results are expected to provide insight into inhalation efficiency of different aerosol types, and the likelihood of deposition in different nasal cavity surfaces.

	\end{abstract}
}	
	
\end{frontmatter}

\section{Introduction}
\label{Introduction}
Nasal airway obstruction can lead to significant impairment in quality-of-life, secondary to a myriad of potential sequelae, including chronic nasal obstruction, mouth breathing, chronic rhinosinusitis, sleep-disordered breathing and obstructive sleep apnoea. As the prevalence of nasal airway obstruction is high, surgery to relieve nasal obstruction is a commonly performed elective procedure. Post-surgery patients have significantly altered anatomy and simulating airflow will improve our understanding of how surgical strategies affect post-surgical airway ventilation, as well as the inhalation dosimetry caused by exposure to airborne pollutants. The removal of anatomical structures in the nose opens up the airway, redistributing the airflow and consequently the particle distribution. 

There have been extensive computational studies investigating different nasal surgeries and their effect on nasal airflow \citep{Xiong2008, Rhee2011, DeWang2012, Frank2013a, Hariri2015, Dayal2016,Maza2019,  Inthavong2019, Siu2020a}, which showed the ability of virtual surgery to guide surgical interventions to alleviate nasal airway obstruction.  Previous CFD studies of middle turbinate resection \citep{Zhao2014, Alam2019, Lee2016} have demonstrated shifts in airflow toward the area of middle turbinate resection and decreased airflow velocity, decreased wall shear stress, and increased local air pressure.

\cite{Chen2012} found that following functional endoscopic sinus surgery, a moderate inspiratory airflow rate and a particle diameter of approximately 10$\upmu$m improved intranasal deposition into the maxillary sinuses. \cite{Frank2013} investigated surgical correction of nasal anatomic deformities on ten patients, and their simulation results showed  posterior particle deposition after surgery increased by 118\%. \cite{Bahmanzadeh2015} showed that sphenoidotomy increased the airflow and particle deposition into the sphenoid sinus, with the highest deposition occurring for resting breathing rate with $10\upmu$m. \cite{Siu2020} studied drug delivery in post-operative sinonasal geometries of patients who underwent a comprehensive functional endoscopic sinus surgery and a modified endoscopic Lothrop procedure and found sinus deposition was more effective with inhalational transport of low-inertia particles outside of the range produced by many standard nasal sprays or nebulizer.

A recent outcome from \cite{Delgaudio2019}’s analysis of patients with middle turbinate resection suggested that the surgery removed the middle turbinate protective function in preventing inhaled allergen particles from depositing on the septum. Anatomical changes in the airway aimed at restoring airway ventilation or increased accessibility to the paranasal sinuses as the primary aims, may in fact be ignoring the secondary effects of persistent inhalation exposure to airborne particulates, such as dust \citep{Tian2007}, dust mites \citep{Zhang2019}, and fibres. Additionally, pollen has been implicated for initiating the onset of allergic rhinitis (such as hay fever) and sometimes has exacerbating asthma \citep{Bousquet2001}.

Modelling of non-spherical particles needs to account for the changes in particle motion due to the different drag coefficients leading to altered trajectories. While the discrete element method and immersed boundary methods can resolve more physics such as rotational degrees-of-freedom and inter-particle interactions, the Lagrangian approach remains popular for its computational efficiency and ability to easily integrate into finite-volume CFD codes. Drag correlations for non-spherical particles have been developed \cite{Holzer2008, Wang2018a,Bagheri2016}.  This is in addition to inclusions of lift and torque components applied to oblate ellipsoidals or fibres \citep{Sanjeevi2018,Zastawny2012,Tian2012, Tian2013}. Th authors used a simplified approach  in earlier work \cite{Inthavong2013}  by using shape factors and an overall drag coefficient to account for tumbling and rotational effects, but this does not allow for the interception deposition mechanism.

This study investigated the changes in deposition of inhaled pollutants in the nasal airways, and the proportion escaping through the nasopharynx towards the lungs. Spheres with aerodynamic equivalent diameters, $d_{ae}$, were used as a reference while pollen particles were investigated. A nasal airway model was used in pre-operative state, followed by a virtual middle turbinectomy surgery (post-operative state) with two constant inhalation flow rates of 5 and 15 L/min. The airway was separated into different regions allowing a detailed comparison of regional deposition changes.

\section{Method}
\subsection{Nasal cavity model}
The nasal cavity geometry was reconstructed from high-quality CT scans of a healthy 25-year-old, Asian female (161cm height, 53kg mass) without any nasal obstructions. The CT data was scanned with a Siemens Dual Source CT Machine with parameters of $0.39 \times 0.39$mm in pixel size, $512 \times 512$ in image dimension, and 0.5mm in slice thickness. The images were imported into 3D Slicer, where the entire paranasal sinus model was determined and exported as a stereolithography (STL) file format for computational modelling. The regions of the sinonasal cavity included the external face, nostrils, nasal cavity, each of the paranasal sinuses and the nasopharynx. Fig \ref{fig:nasalStructure} demonstrates the surface partitioning of the pre-operative model highlighting the local regions used to record deposition values. Each image shows outer walls removed sequentially with the final image depicting the middle turbinate as red: anterior head of the middle turbinate; green: lateral side; and yellow:medial. These regions are removed during the virtual middle turbinectomy resection which was performed in Slicer to create a `\textit{post-operative}' model. 

The segmented models were imported into the commercial CFD software Ansys-Fluent 2020R1 and meshed with polyhedral cells and six prism layers applied in near-wall regions. The advantage of poly-hexcore meshing is that it uses fewer elements, approximate $3.5\times$ fewer than tetrahedral meshing (with the same size functions) while achieving negligibly different results. The mesh strategy was based on a size functions evaluated in an earlier study in  \citep{Inthavong2018} which found that a minimum scoped mesh cell size of 0.45 mm in the nasal cavity region but a much larger, coarse mesh in the outer ambient air region was optimal. During our mesh independence analysis (Supplementary Material) we evaluated the change in velocities in coronal planes resulting in an optimised mesh of 1.58 million cells for the pre-op model and 1.75 million cells for the post-op model. 

Fig \ref{fig:mesh} shows the internal volume mesh taken in the sagittal planes of the left and right cavities. The zoomed view of a posterior near wall region depicts the six prism layers, with polyhedrals  connecting to the hex-core cells to make up the polyhex-core mesh design. The sagittal planes are taken at the same locations which highlights the opened airway caused by surgery in the Post-Op models. The difference in the opened airway is most significant in the right cavity.

\subsection{Fluid particle simulation}
Two constant flow rates were used (5 and 15 L/min), corresponding to shallow and normal resting breathing rates based on a laminar, incompressible flow model. The laminar model provides better overall performance for the flowrates used where the literature \citep{van2021pressure,li2017computational} have demonstrated laminar dominant flow behaviour at flow rates of 15 L/min or less \citep{hahn1993velocity,doorly2008mechanics}.

The flow equations describing the conservation for mass and momentum are expressed as:
\begin{align}
	\frac{\partial u_i}{\partial x_i} &= 0 \\
	\frac{\partial u_j u_i}{\partial x_j} &= -\frac{1}{\rho}\frac{\partial p}{\partial x_i}+\frac{\partial}{\partial x_j}\left( \mu \frac{\partial u_i}{\partial x_j} \right )
\end{align}
where $ u_i $ is the flow velocity vector and $ p $ is the pressure, and $\mu$  is the molecular viscosity. The equations were discretized using the finite volume approach, with upwind second order schemes for the pressure, and momentum equations. The pressure-velocity coupling used the 'Coupled' scheme and the solution was iterated through the 'pseudo transient' option and a time scale factor of 0.01 s. The pseudo-transient method improves stability and speeds up convergence through a form of implicit relaxation (as opposed to explicit relaxation). The convergence criteria was set to 1e-5 where the residual values were calculated based on local scaling.

In this study, only the gravity and the drag force were considered for tracking the aerodynamic sphere and pollen particles. The particle sizes of interest were in the micron size range, and therefore forces that affect submicron particles such as Brownian, and lift forces were excluded as they are negligible for micron particles. The particle equation of motion is given by
\begin{equation}
\frac{\mathrm{d}u_p}{\mathrm{d}t}=F_d+\frac{g\left ( \rho_p-\rho_g \right )}{\rho_p}
\end{equation}

where the subscripts $p$ and $g$ represent particle and gas (air) phases, $\rho$ is the density $g$ is the gravity term and $F_d$ is the drag force per unit mass. 
\begin{equation}
F_d = \frac{18\mu}{\rho_p d_p^2}\frac{C_d\mathrm{Re}}{24}(u_g-u_p)=\frac{(u_g-u_p)}{\tau}
\end{equation}
where $\tau$ is the particle relaxation time, and $d_p$ is taken as the volume equivalent diameter $d_v$. For normalising particles with different densities the aerodynamic  diameter is related by
\begin{equation}
d_{ae}=d_g \sqrt{\rho_p/\rho_0}
\end{equation} 
where $d_g$ is the geometric particle diameter, $\rho_p$ is the particle density, and $\rho_0$ is the particle unit density equal to 1 g/cm$^3$. For the spherical particle,  the drag coefficient $C_d$  is taken from \cite{Morsi1972} defined by
\begin{equation}
C_d  = a_1 + \frac{a_2}{\mathrm{Re}_p}+ \frac{a_3}{\mathrm{Re}_p}
\label{eqn:morsi}
\end{equation}

where the $a_1$, $a_2$, and $a_3$ are empirical constants for smooth spherical particles over different ranges of particle Reynolds number.

For pollen, an assumption of a non-spherical particle shape based on the experimental measurements of \cite{Tran-Cong2004} was used to define the non-spherical shapes through an aggregation of particles clustered together. Fig \ref{fig:pollen} shows scanned electron microscope images of pollen reported in \cite{Suarez-Cervera2008} and from Dartmouth College. The rough surface non-spherical particle assumption was used following \cite{Tran-Cong2004}'s technique (Fig \ref{fig:pollen}c). Two equivalent diameters, volume equivalent diameter $d_n$, and surface equivalent diameter $d_a$ and a shape factor called the ‘degree of circularity’ \cite{Wadell1933} were used for the drag correlation given as,
\begin{equation}
d_n = \sqrt[3]{6V/\pi}; \hspace{2cm}  d_a = \sqrt{4 A_{proj}/\pi}; \hspace{2cm} c = \pi\frac{d_A}{P_{proj}}
\end{equation}
where $V$ is the particle volume, and $A_{proj}$ is the projected area of the sphere. The circularity $c$ is defined as
where $P_{proj}$ is the projected perimeter of the particle in its direction of motion. The empirically defined correlation for the drag coefficient for the pollen is given as
\begin{equation}
C_D=\frac{24}{\textrm{Re}_p}\frac{d_a}{d_n}\left [ 1+ \frac{0.15}{\sqrt{c}} \left ( \frac{d_a}{d_n}\textrm{Re}_p \right )^{0.687}\right ]+\frac{0.42\left ( \frac{d_a}{d_n} \right )^2}{\sqrt{c}\left [  1+42500 \left ( \frac{d_a}{d_n} \textrm{Re}_p\right )^{-1.16}\right ]}
\label{eqn:trancong}
\end{equation}

This correlation was implemented into Ansys-Fluent via a user-defined function allowing a customised drag correlation and has been used in a previous study \cite{Inthavong2008}.  The proposed correlation is valid in the ranges of variables
$0.15 < Re < 1500$; $0.80 < d_a/d_n < 1.50$; and $0.4 < c < 1.0$. Particles were individually tracked through the nasal cavity using a one-way coupled Lagrangian Discrete Phase Model, where the airflow field was first simulated and subsequently trajectories of each particle determined by integrating the particle force balance equation while accounting for gravity and drag. A particle number independence was evaluated which showed results began converging with less than 5\% change in deposition results from 15,000 to 20,000 particles. To ensure particle number indpendence, approximatly 30,000 particles were introduced for each particle diameter tested (e.g. $n=16$ diameters $\times 30,000$ particles). The initial particle locations to represent outside air exposure under the influence of inhalation flow rates was set by distributing the particles uniformly on a spherical dome 3~cm from the face based on the breathing region distance \citep{Shang2015, Zhu2005}. The particles were set with a zero velocity, and the spherical dome distance from the nostrils allowed the particles to be naturally inhaled under the influence of the inhalation flow rates. 

The morphology of pollen comes in various shapes, diameters and densities, which is subjected to different aerodynamic properties. For example, \cite{Crowder2002} reported geometric diameters ranging from $16-27\upmu$m, \cite{Kelly2002pollen} found a range between $10-60\upmu$m. A density of 550kg/m$^3$ for pollen was adopted based on \cite{Crowder2002}. The pollen shape exhibited spiny protrusions that arise from a spherical surface base (see Fig \ref{fig:pollen}). The geometric diameters for the spherical and pollen particles used in this study ranged from 4 to 70 microns (4, 6, 8, 10, 12, 14, 16, 18, 20, 25, 30, 35, 40, 50, 60, 70).

\section{Results}
\subsection{Flow Field}
Velocity contours on ten coronal planes from the anterior, main nasal passage and the posterior regions are shown in Fig \ref{fig:planes}. The highest velocities were found in the anterior planes, which exhibited the smallest cross-sectional areas. The white to red colours in the contours depict the main bulk flow paths which were located in the medial regions of the planes as the airflow is directed superiorly from the nostrils. The effect of surgery is evident in planes $y4$ to $y8$ where the removal of the turbinate bone creates a large open space in the superior half of the main nasal passage. The flow becomes more disturbed and has greater penetration into the most superior locations of the cavity. 

Three sagittal planes were taken in both the left and right cavities (for a total of six planes), and their velocity contours are shown in Fig \ref{fig:sagCont}. The middle turbinate is located by a yellow surface which disrupts the main flow paths in the Pre-Op model. In contrast, in the Post-Op model the airflow moves freely towards the ceiling of the nasal cavity (shown in lateral, and middle planes). In the septal sides of both left and right cavities, the planes show the septal wall protruding into the anterior airway, and this is confirmed in plane $y3$ of the anterior coronal planes from Fig \ref{fig:planes}. It is expected that inhaled particles will accelerate with the flow field directed at an angle shown in the velocity contours in the sagittal planes. Furthermore, the particles have enhanced opportunity to deposit on anteriorly protruding anterior septal wall, and on the middle turbinate in the Pre-Op model, and nasal cavity ceiling on the Post-Op model.

\subsection{Inhalation Efficiency}
The inhaled fraction of particles released from an upstream radial position of 3-cm from the nostrils was determined, shown in Fig \ref{fig:inhalation}. The radial distance selected followed the work from the face was based on the breathing influence and region \citep{Shang2015, Zhu2005}. The inhalation fraction differs from the inhaled efficiency measure (also known as aspiration efficiency), where the latter is determined from the ratio of aspirated particle concentration to the ambient freestream particle concentration \citep{Kennedy2002, Dai2006, Li2016, Tao2020, Dai2006, KingSe2010}. 

The inhaled fraction of pollen demonstrated greater opportunities of inhalation compared to aerodynamic spheres. The spherical particles were treated as aerodynamic diameters with $\rho_p = 1000$kg/m$^3$ which is $1.82 \times $ greater than pollen which contributes to the spherical particle inertia, mass, and its sedimentation away from the inhaled air. For the same diameters, the inhaled fraction of the lighter pollen particles is significantly greater than the spheres. When the pollen particle diameters are normalised by it's aerodynamic diameter (Fig \ref{fig:inhalation}b) it's inhalation fraction curve moves closer to the spherical particle, but remains greater for the same aerodynamic diameter.

A secondary influence on the aerodynamic flight of the pollen and spherical particles is the difference in drag coefficient shown in Fig \ref{fig:inhalation}c for Re number range of $10^{-1}$ to $10^3$. The pollen experiences greater drag force suggesting it adapts to the inhaled air easier than a spherical particle. The combined effects of lower density and a higher drag coefficient in the pollen led to greater chances of inhalation than in an aerodynamic spherical particle. The multi-parameters that affect the inhalation fraction can be collapsed into a single term which we define as the `\textit{inhalation parameter}' as:
\begin{equation}
	\mathrm{IP} = d_{ae}^2Q^{-b}
\end{equation}
where $b$ was a variable tuned to curve fit the deposition data, which was $b=0.825$. This combines the inhalation rate $Q$ in units of [cm$^3$/s], aerodynamic equivalent diameter $d_{ae}$ in units of $\upmu$m to conveniently capture the influence of inhalation rate and particle density.  The data collapsed onto a single trendline for either particle type allowing curve fitted correlations to predict the inhalation fraction from exposure to particles released 3~cm from the nostrils.

\subsection{Local Surface Deposition}
The local deposition fraction for spherical and pollen particles on individual surfaces are shown on Fig \ref{fig:depPlot}. In general, the particle diameters achieving peak deposition shifted to the right (increased diameter sizes) for the right nasal cavity compared with the left, suggesting the anatomical differences in the cavities led to different deposition behaviour. In the left cavity, deposition was found predominantly on the nasal septum, and the effects of surgery were generally slightly higher. In the right nasal cavity, deposition was mainly found on the anterior middle turbinate in Pre-Op. 

After removal of the middle turbinates, the particles travelled further into the nasal airway and deposited on the upper lateral walls. Overall deposition decreased in the Post-Op model at 5L/min, but was relatively constant at 15L/min. Deposition on the nasal septum was relatively similar between Pre-Op and Post-Op. There was increased deposition for the higher flow rate of 15L/min with more particles depositing in the anterior regions. The right cavity of the Post-Op model also exhibited increased deposition in the paranasal sinuses, which was negligible in the Pre-Op models.

The deposition sites on the nasal cavity are shown in Fig \ref{fig:depSites} where the particles are coloured by particle diameter). There is a consistent concentration of particles depositing on the anterior nasal septum that protrudes into the airway (circled once in the Fig \ref{fig:depSites}c). At 15L/min a second local concentration site is found at the vestibule. These two regions account for the high inertial particles (black to red coloured diameters) impacting on surfaces directly aligned to the inhalation path. At 5L/min more inhaled particles penetrate the anterior nasal valve, and when the airflow slows in the main nasal passage, the particles succumb to gravitational sedimentation and deposition is found along the nasal floor. 

Deposition in the main nasal passage appears to scatter across all surfaces randomly. Fig \ref{fig:depSeptum} extracted deposition on the nasal septum before and after surgery with the middle turbinates were included in Pre-Op model. The head of the anterior middle turbinate is the frontline region for contact with the inhaled air and particles, and it is expected to receive high deposition. The lateral, and medial middle turbinate regions squeeze into the airspace and deposition in these regions arise from changes in the flow streamlines (i.e. changes in flow direction). The Post-Op images show the effect of middle turbinectomy resection, but there are no consistent changes among all cases. In general, however, there is an increase in deposition in the superior half of the nasal septum, attributed to the more open passageway allowing more flow movement the nasal ceiling.

\subsection{Deposition Trends}
Fig \ref{fig:swarm} collects all the deposition data into a single `swarm' plot (similar to a box plot). The values spread laterally for repeating deposition fraction values; thus, a wide section of the plot suggests a common occurrence of the data. The height represents the occurrence of high deposition fractions in the specific surface. The results for the Pre-Op model shows that the highest deposition occurs in the nasal septum ($>0.1$), and this is only achieved for inhalation rates of 15L/min (red markers only). The next highest deposition occurs in the anterior middle turbinate (up to a fraction of 0.3). In the Anterior, and to a lesser extent the Lateral-Upper regions, the higher flow rate enhanced their deposition. In the Post-Op model (Fig \ref{fig:swarm}b)the middle turbinate regions (MT-anterior, MT-lateral, MT-medial) are removed, and there is an increase in the Lateral-Upper, and Sinuses. At the same time, there is very low deposition in the Lateral-Lower, and Turbinate (superior, and inferior) regions.

The change in deposition fraction due to surgery is obtained by subtracting Pre-Op values from Post-Op values shown in Fig \ref{fig:swarm}c. The results indicate that middle turbinate resection in this patient generally increased deposition in the Lateral-Upper, Sinuses, and the Septum. The highest increases in deposition occurred for either particle inhaled at 15L/min. Increased deposition between 0.0 to 0.03 in the Septum and Sinuses were produced from a combination of particle type and inhalation rate. A reduction in the deposition was found in all surfaces except the sinuses, and this was predominantly caused by inhalation rates of 5L/min. Fig \ref{fig:swarm}d shows the same change in deposition fraction due to surgery, but this time the markers are coloured by nasal cavity side. This highlighted the left nasal cavity as the main contributor for deposition increases in the Sinuses, and Lateral-Upper. In contrast, the right nasal cavity demonstrated lower values of deposition decreases in the Lateral-Upper, Lateral-Lower, and Turbinate surfaces.

\section{Discussion}
Pollen particles exhibited higher drag coefficients, and a lower particle density compared to aerodynamic equivalent spheres. This suggests that pollen has greater mobility in its aerodynamic flight and greater potential to penetrate the nasal cavity. Middle turbinate resection in this study created a larger open space in the superior half of the main nasal passage where the inhaled air was directed into, thus transporting inhaled particles to the space. At 5L/min, smaller particles penetrated deeper into the nasal cavity and leaving through the nasopharynx exit, leading a decrease in nasal deposition.  At 15L/min, the inertial properties of the particles are enhanced threefold, and this led to increased anterior deposition, particularly the protruding nasal septum near the nasal valve. 

The middle turbinate directly faces the oncoming flow and bifurcates the air into lateral and septal streams. This suggests that the propensity for deposition onto either surface is sensitive to the orientation of the anterior middle turbinate relative to the oncoming air. The left cavity exhibited very high deposition on the nasal septum before surgery, whereas the right cavity exhibited more evenly distributed deposition across different surfaces but predominantly across the anterior, middle turbinate, and the septum. The larger open space produced from the surgery increased accessibility to the superior half of the main nasal passage, and this was reflected by the increase in deposition in the `Lateral-Upper' region, and this was found only in the left nasal cavity at 15L/min inhalation flow rate. There was an increase in the nasal septum, but to a lesser extent, however, the changes may be masked by the initially high deposition in the anterior nasal septum in the Pre-Op model. 

CFD analysis demonstrates that, in the unoperated normal model, a large number of particles deposit on the middle turbinate, particularly on the anterior edge and lateral surface of the middle turbinate, along with the septum. Following resection of the middle turbinatess, particle deposition increases at the septum and at the upper lateral nasal walls. These deposition trends demonstrates that middle turbinate excision leads to increased allergen deposition on the septum, confirming the clinical suspicion that the middle turbinate plays a protective role in preventing allergen impaction on the septum. In patients with aspirin exacerbated respiratory disease (AERD), surgical strategies aimed at preserving the middle turbinate should be considered.

\section{Conclusion}
An inhalation parameter was introduced to account for the multi-parameter scenario of inhaled particles from the surrounding air, given as $d^2_{ae}Q^{-0.825}$. The drag coefficient for pollen and its lighter density led to greater inhalation efficiency compared to spherical particles at 1000kg/m$^3$. Correlations to curve fit the pollen, and spherical particle was produced to predict inhalation fractions. Deposition on local surfaces was obtained for Pre-Op and Post-Op models. The effect of surgery showed an increase in deposition in the Post-Op at 15L/min but a decrease at 5L/min. Middle turbinate resection created a more open space in the superior half of the main nasal passage, increasing accessibility to the region. The increase in deposition in the Post-Op model occurred in the left cavity in the Lateral-Upper area. There was also an increase on the nasal septum, but this value may be masked by the initially high deposition already found in the Pre-Op model. It is expected that surgical planning of middle turbinate resection should consider the implications of increased allergen deposition on the septum which can lead to secondary sequela. This is particularly critical for patients with aspirin exacerbated respiratory disease (AERD). Future work will investigate more population samples/models, and the influence of transient inhalation breathing flow.

\section{Acknowledgements}
The authors gratefully acknowledge the financial support provided by the Garnett Passe and Rodney Williams Foundation Conjoint Grant 2019-22.

\small
{\setstretch{1.0}
\bibliography{pollen}

\normalsize
\newgeometry{margin=2.0cm}

\clearpage
\newgeometry{margin=1.5cm}

\clearpage
\begin{figure}
	\centering
	\includegraphics[width=0.9\linewidth]{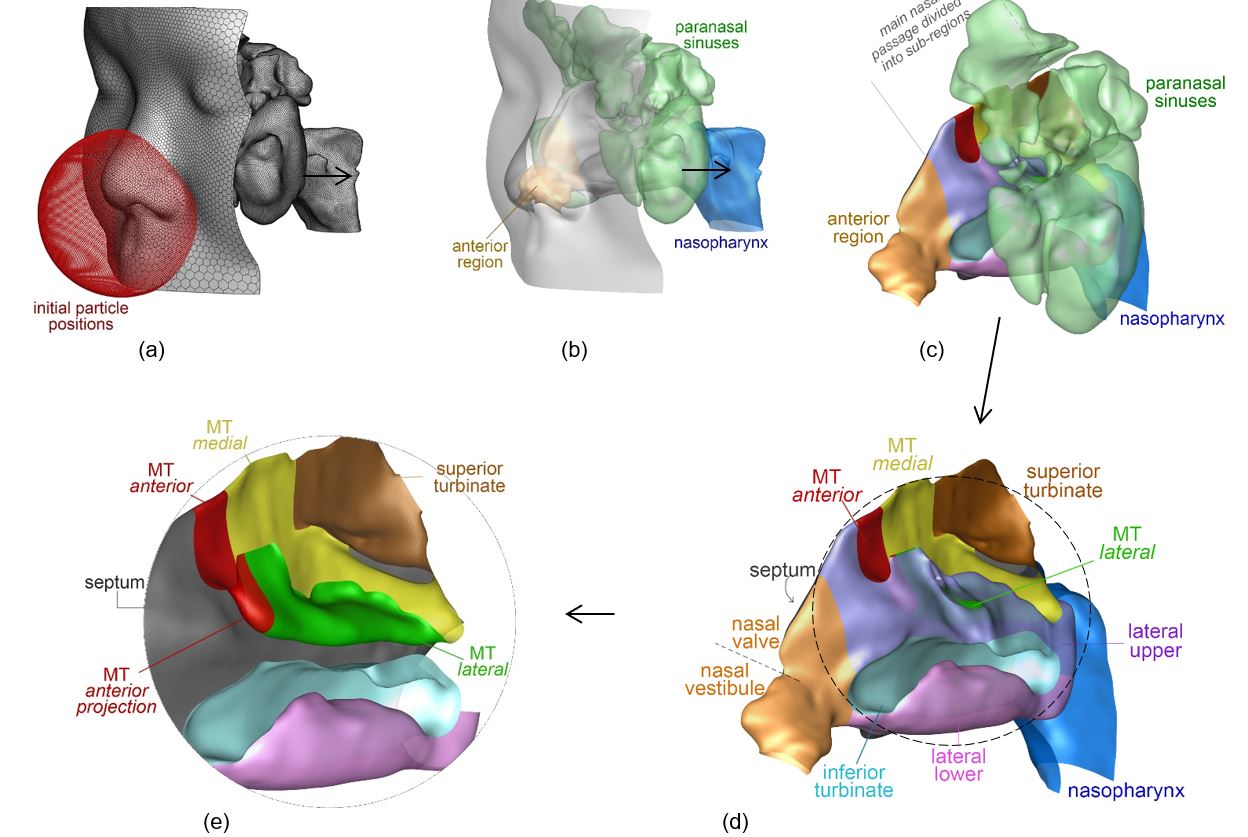}
	\caption{Computational model depicting (a) the surface mesh, and the initial particle positions (labelled in red) used for inhalation exposure analysis; (b) the face, anterior region (orange) which consists of the nasal vesitbule, and nasal valve, paranasal sinuses (green), nasopharynx (blue); (c) the face removed; (d) the main nasal passage with individual sub-regions; followed by (e) a zoomed view of the middle turbinates with the outer lateral walls removed.}
	\label{fig:nasalStructure}
\end{figure}

\clearpage
\begin{figure}
	\centering
	\includegraphics[width=0.95\linewidth]{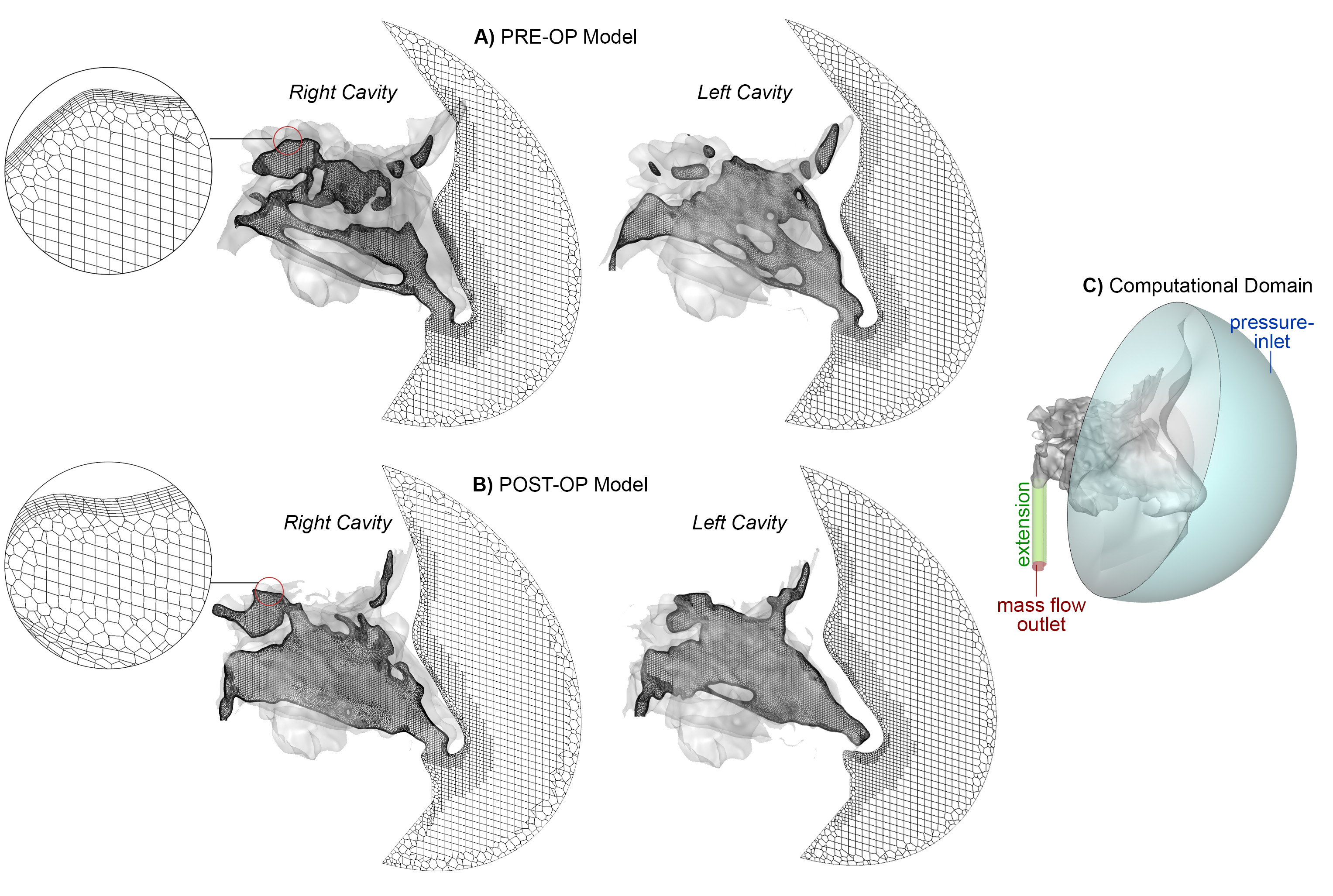}
	\caption{Internal volume mesh shown on sagittal planes in the left and right cavities for the (A) Pre-Op and (B) Post-Op models. A zoomed vieew of the mesh near the wall depicts the prism layers which are connected to the hex-core mesh elements through polyhedral cells. (C) The computational domain is shown with the imposed boundary conditions. An extension was added at the nasopharynx to prevent backflow occurring at the outlet.}
	\label{fig:mesh}
\end{figure}

\clearpage
\begin{figure}
	\centering
	\includegraphics[width=0.7\linewidth]{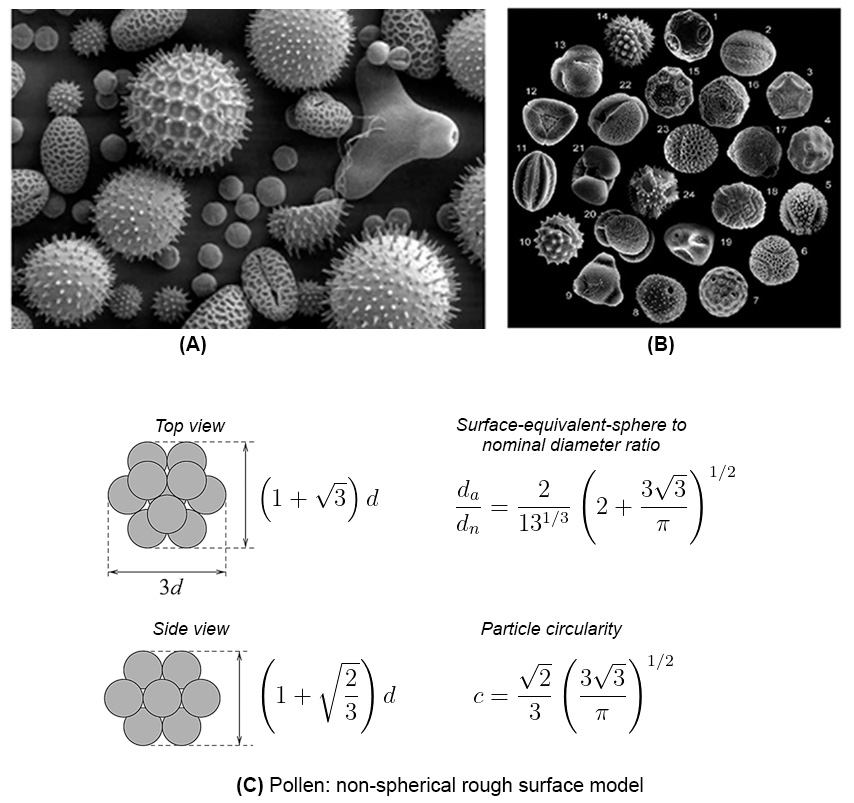}
	\caption{Scanning electron microscope images of pollen grains reproduced from: (a) Louisa Howard and Charles Daghlian of Dartmouth College – . Zeiss DSM 962 SEM, Rippel Electron Microscope Facility; and (b) \cite{Suarez-Cervera2008} demonstrating pollen morphological diversity .}
	\label{fig:pollen}
\end{figure}

\clearpage
\begin{figure}
	\centering
	\includegraphics[width=0.9\linewidth]{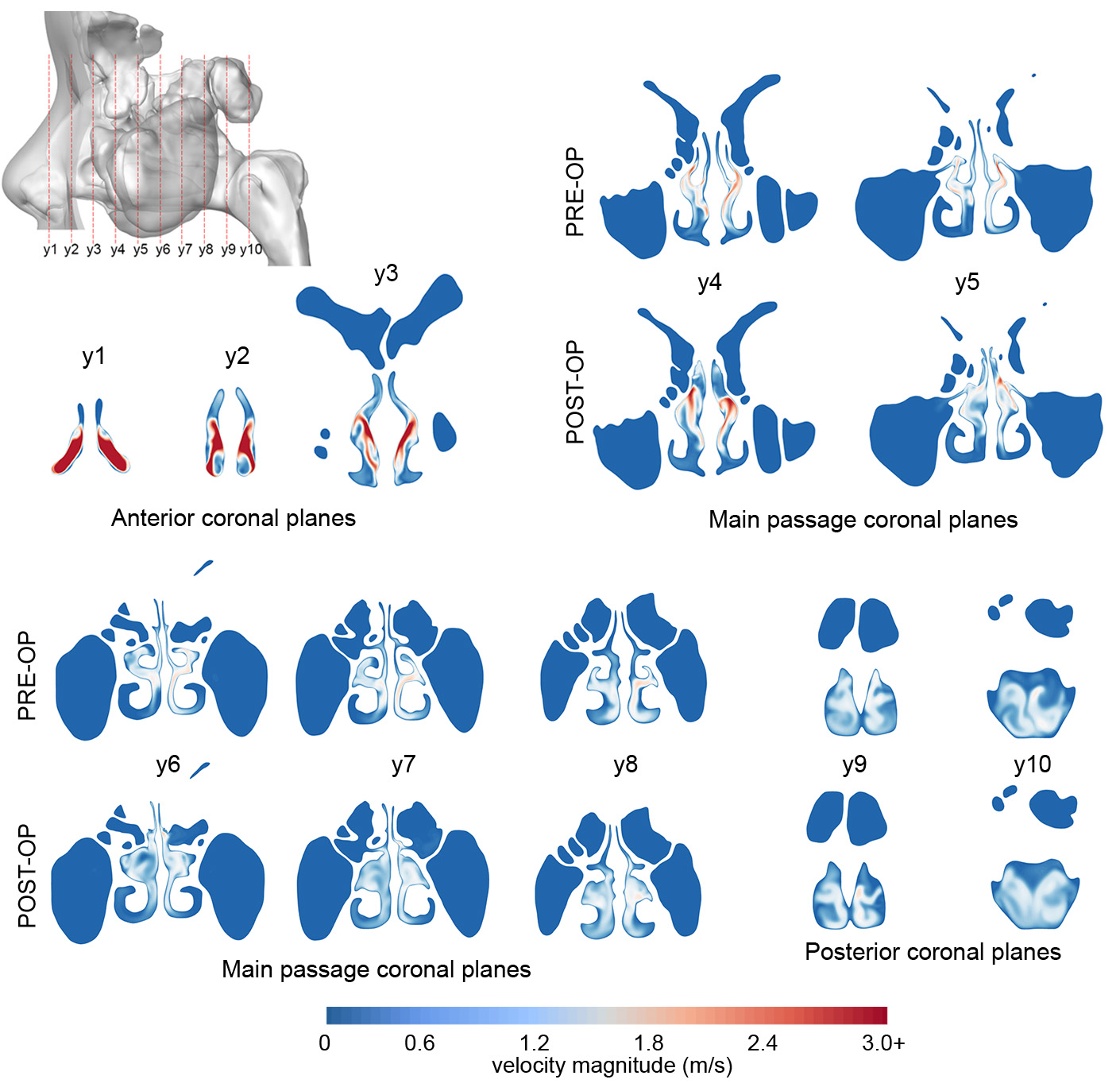}
	\caption{Ten coronal slices (viewed from the face) taken in the anterior, main nasal passage, and the posterior regions. The anterior and posterior slices were identical in the Pre-Op and Post-Op models, but differed in the main nasal passage, where the middle turbinate resection took place. Velocity contours for the anterior planes (y1 to y3) were identical and therefore only one set is shown. The main passage (y3 to y8) and posterior planes (y9, y10) demonstrated different velocity flow fields. The peak velocity was 4.1m/s in the anterior coronal planes. The contour legend was capped at 3m/s to provide better contrast in colours in the main and posterior planes.}
	\label{fig:planes}
\end{figure}

\clearpage
\begin{figure}
	\begin{subfigure}[b]{1\textwidth}
		\centering
		\includegraphics[width=0.35\linewidth]{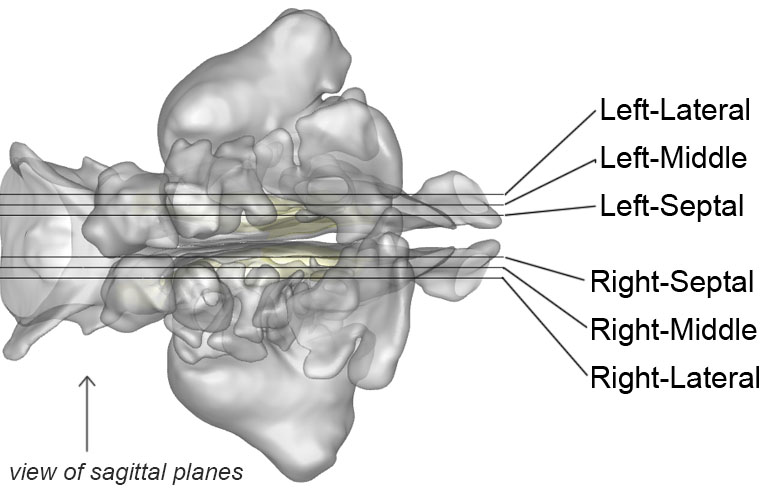}
		\caption{Sagittal plane locations}
		\vspace*{2mm}
	\end{subfigure}
	~
	\begin{subfigure}[b]{1\textwidth}
		\centering
		\includegraphics[width=0.65\linewidth]{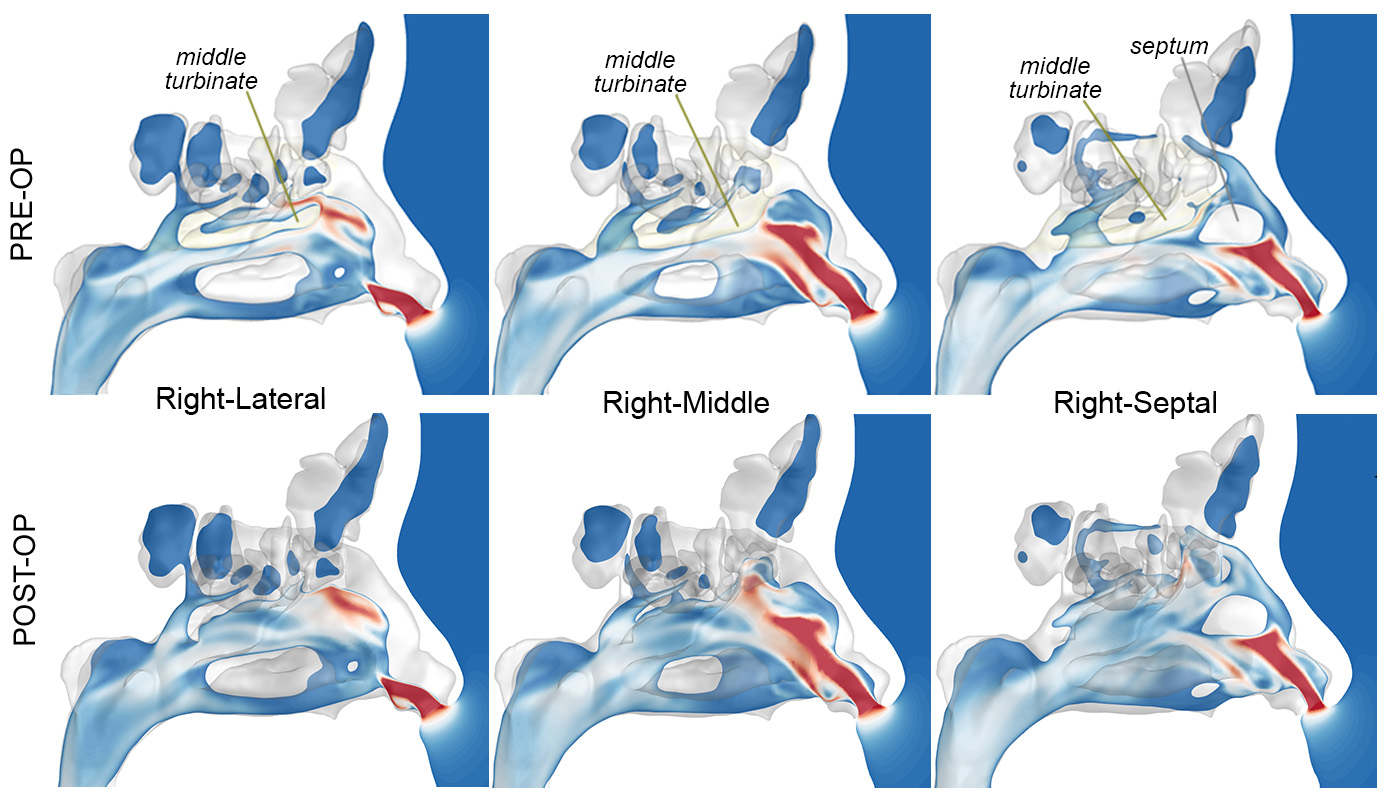}
		\caption{Sagittal planes in the right nasal cavity}
		\vspace*{2mm}
	\end{subfigure}
	~
	\centering
	\begin{subfigure}[b]{1\textwidth}
		\centering
		\includegraphics[width=0.65\linewidth]{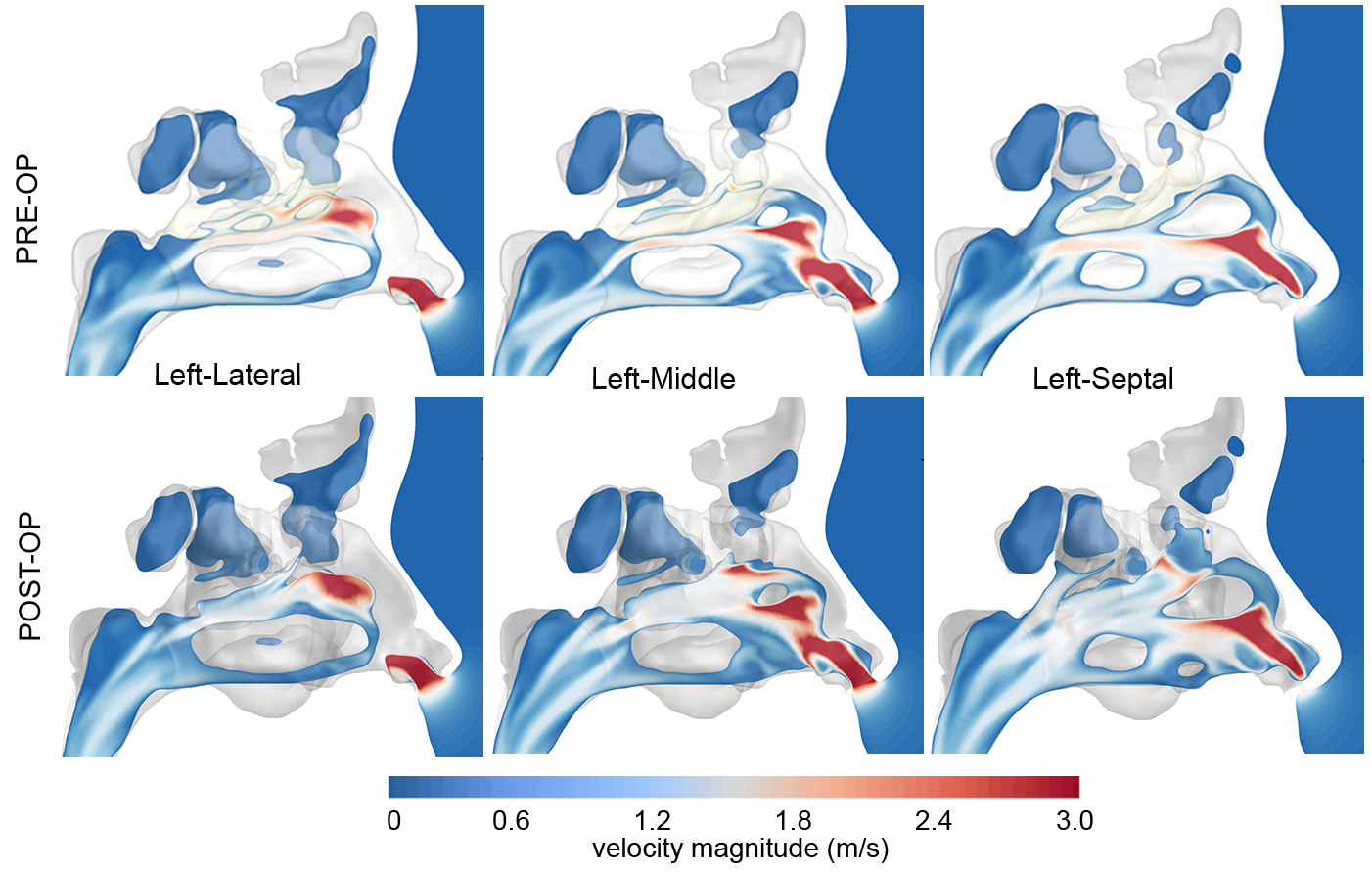}
		\caption{Sagittal planes in the left nasal cavity}
		\vspace*{2mm}
	\end{subfigure}

	\caption{(a) Top view, looking down at the nasal cavity showing sagittal planes in the left and right nasal cavities. (b) Velocity contours in the right nasal cavity, and; (c) velocity contours in the left nasal airway for Pre-Op and Post-Op models.} 
	\label{fig:sagCont}
\end{figure}

\clearpage
\begin{figure}
	\centering
	\begin{subfigure}[b]{0.475\textwidth}
		\includegraphics[width=1\linewidth]{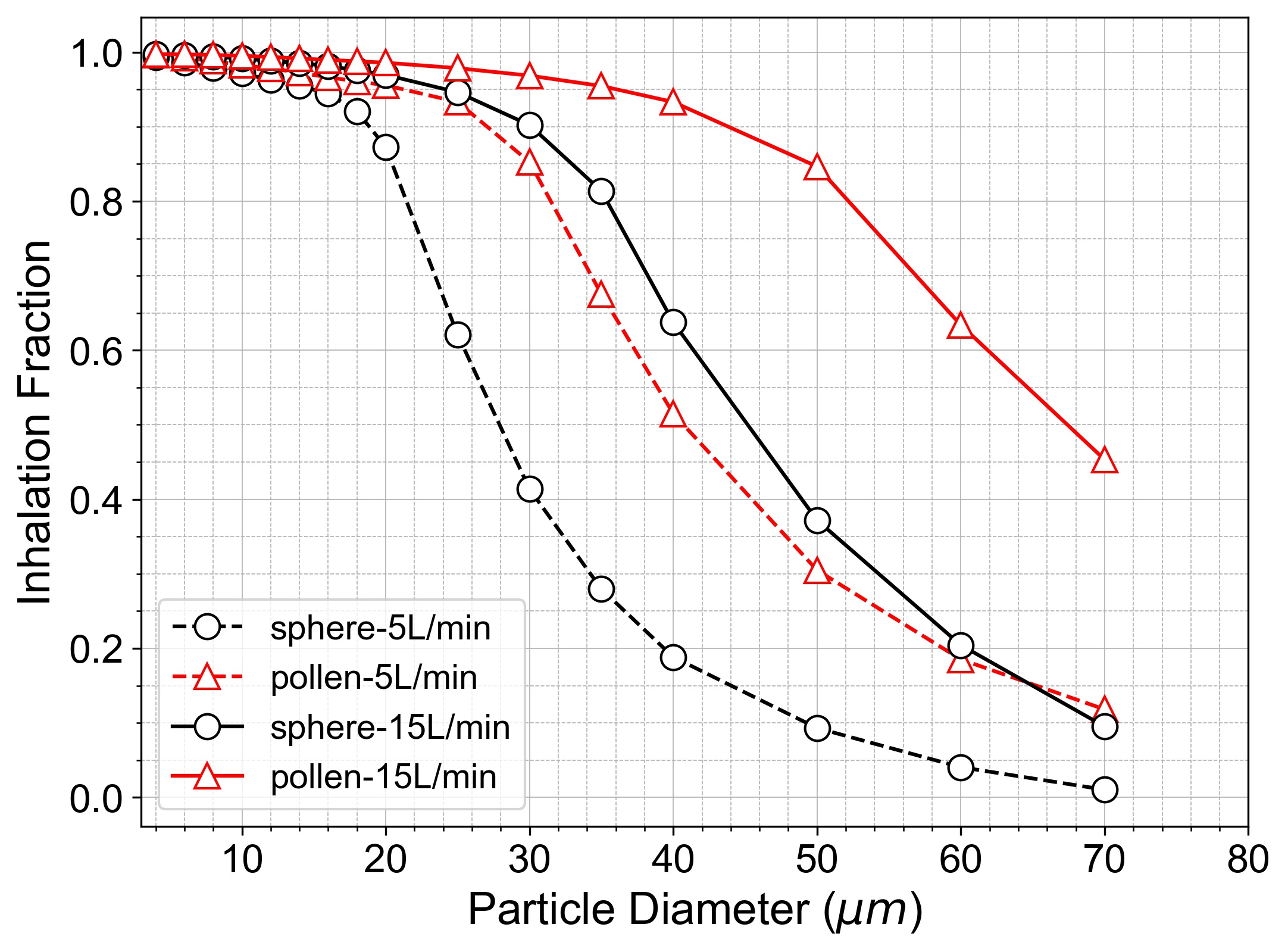}
		\caption{Inhalation fraction for sphere and pollen}
		\vspace*{2mm}
	\end{subfigure}
~
	\begin{subfigure}[b]{0.475\textwidth}
	\includegraphics[width=1\linewidth]{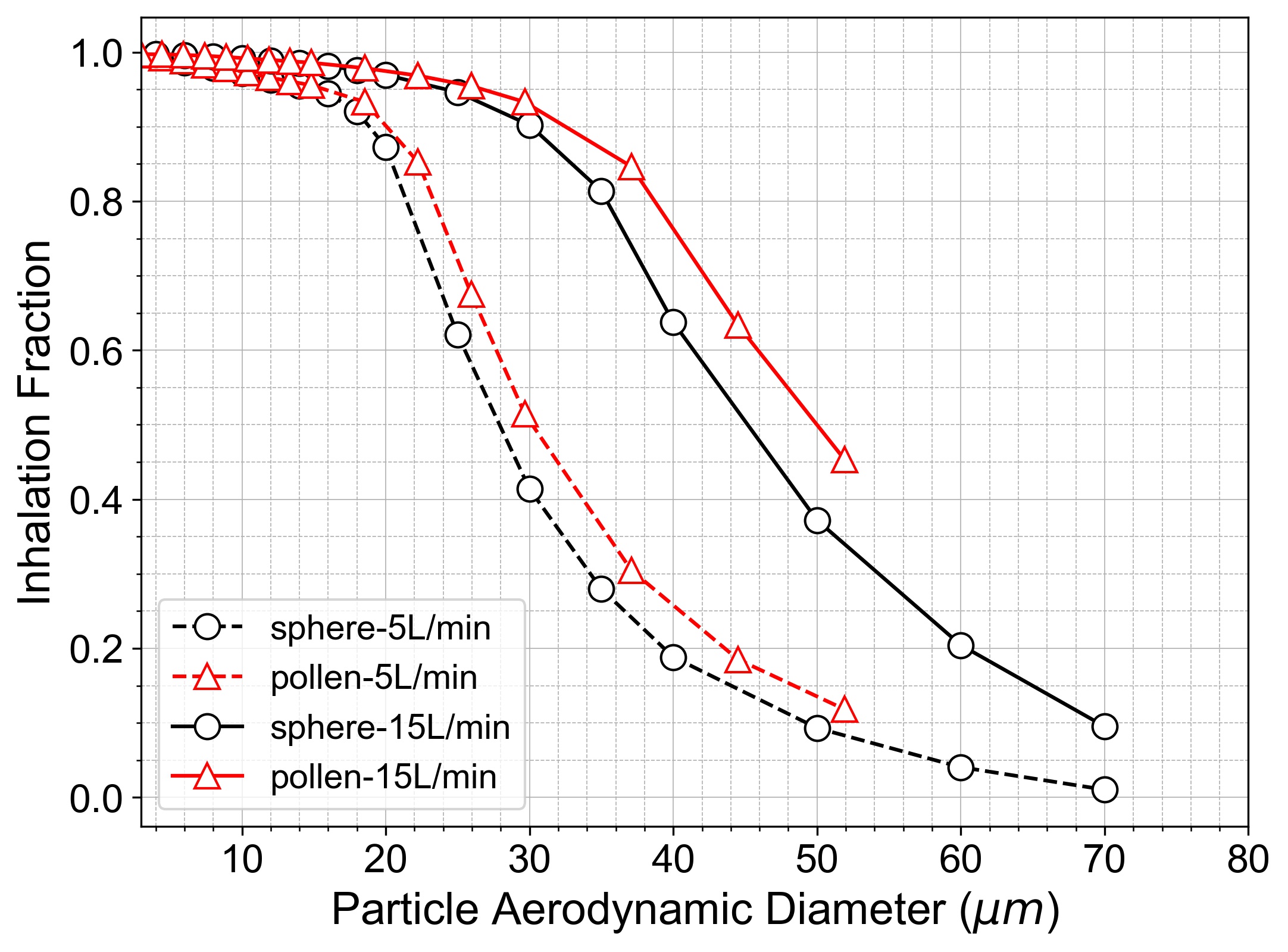}
	\caption{Inhalation fraction normalised by aerodynamic diameter}
	\vspace*{2mm}
\end{subfigure}
~

	\centering
\begin{subfigure}[b]{0.475\textwidth}
	\includegraphics[width=1\linewidth]{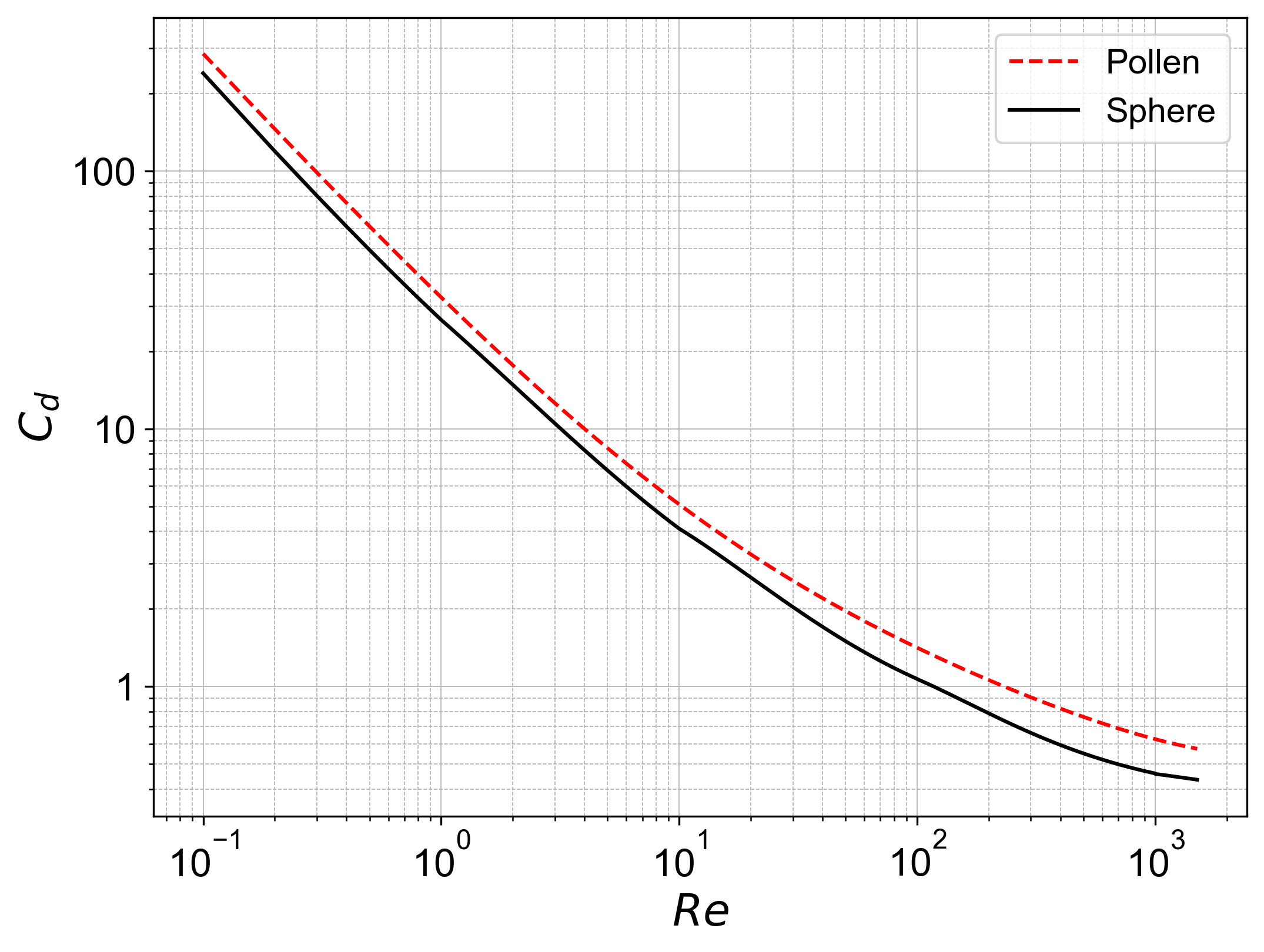}
	\caption{Drag coefficient for a sphere and pollen particle}
	\vspace*{2mm}
\end{subfigure}
~
\begin{subfigure}[b]{0.475\textwidth}
	\includegraphics[width=1\linewidth]{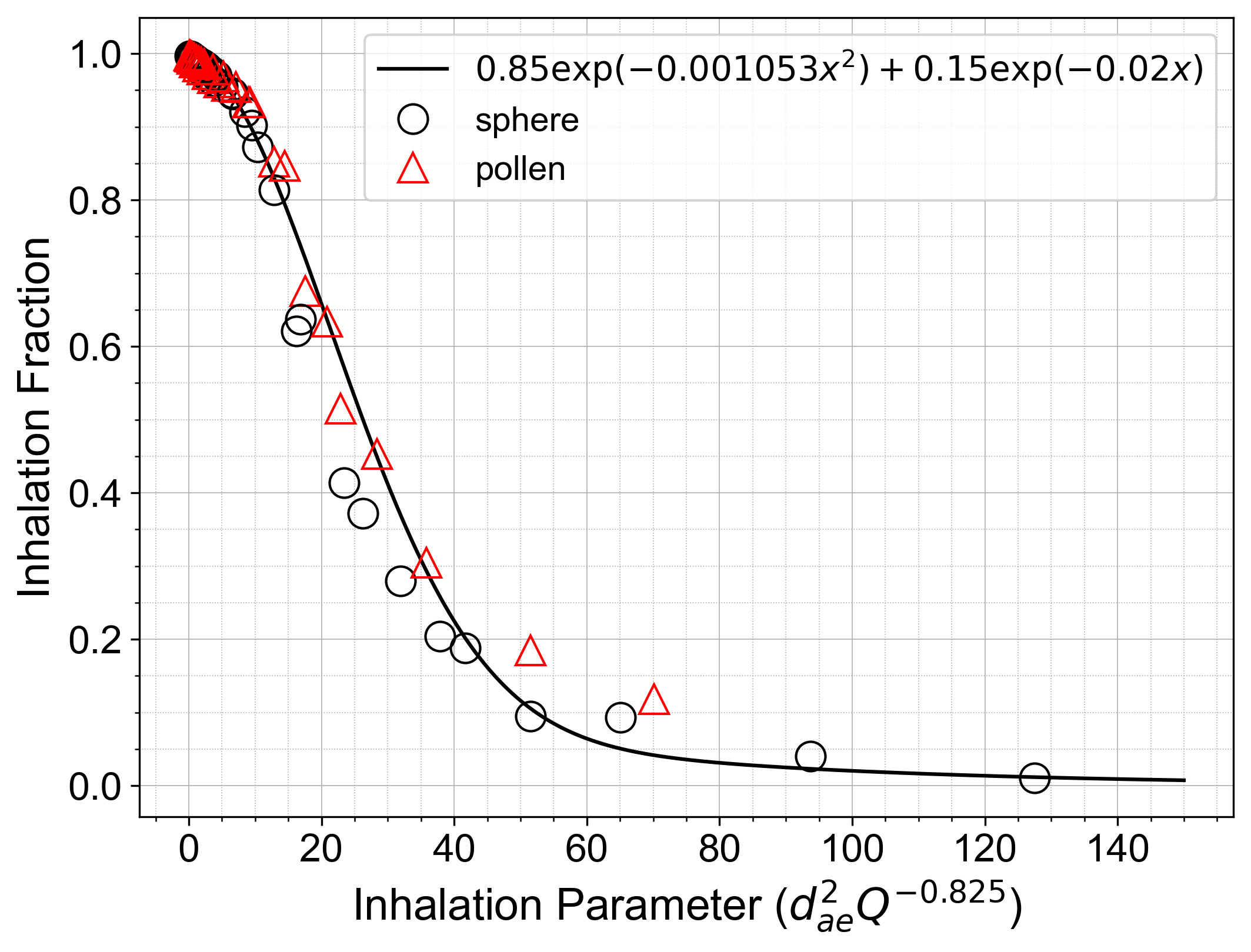}
	\caption{Curve fit of inhalation fraction for sphere, pollen.}
	\vspace*{2mm}
\end{subfigure}
	
	\caption{(a) Inhalation fraction for sphere and pollen particles at 5L/min and 15L/min inhalation rates. (b) Inhalation fraction for sphere and pollen normalised by aerodynamic diameter. (c) Drag coefficient values for pollen (Eqn \ref{eqn:morsi}) and sphere (Eqn \ref{eqn:morsi}). (d) Inhalation fraction as a function of an inhalation parameter for pollen and spherical particles released at 3-cm radius from nostril inlets. The curve fit correlation provides a method for normalising the inhaled fraction $IF = 0.85 \exp(-0.001053x^2)+0.15\exp(-0.02x)$} 
	\label{fig:inhalation}
\end{figure}

\clearpage
\newgeometry{margin=0.75cm}

\clearpage
\begin{figure}
	\centering
	\includegraphics[width=1\linewidth]{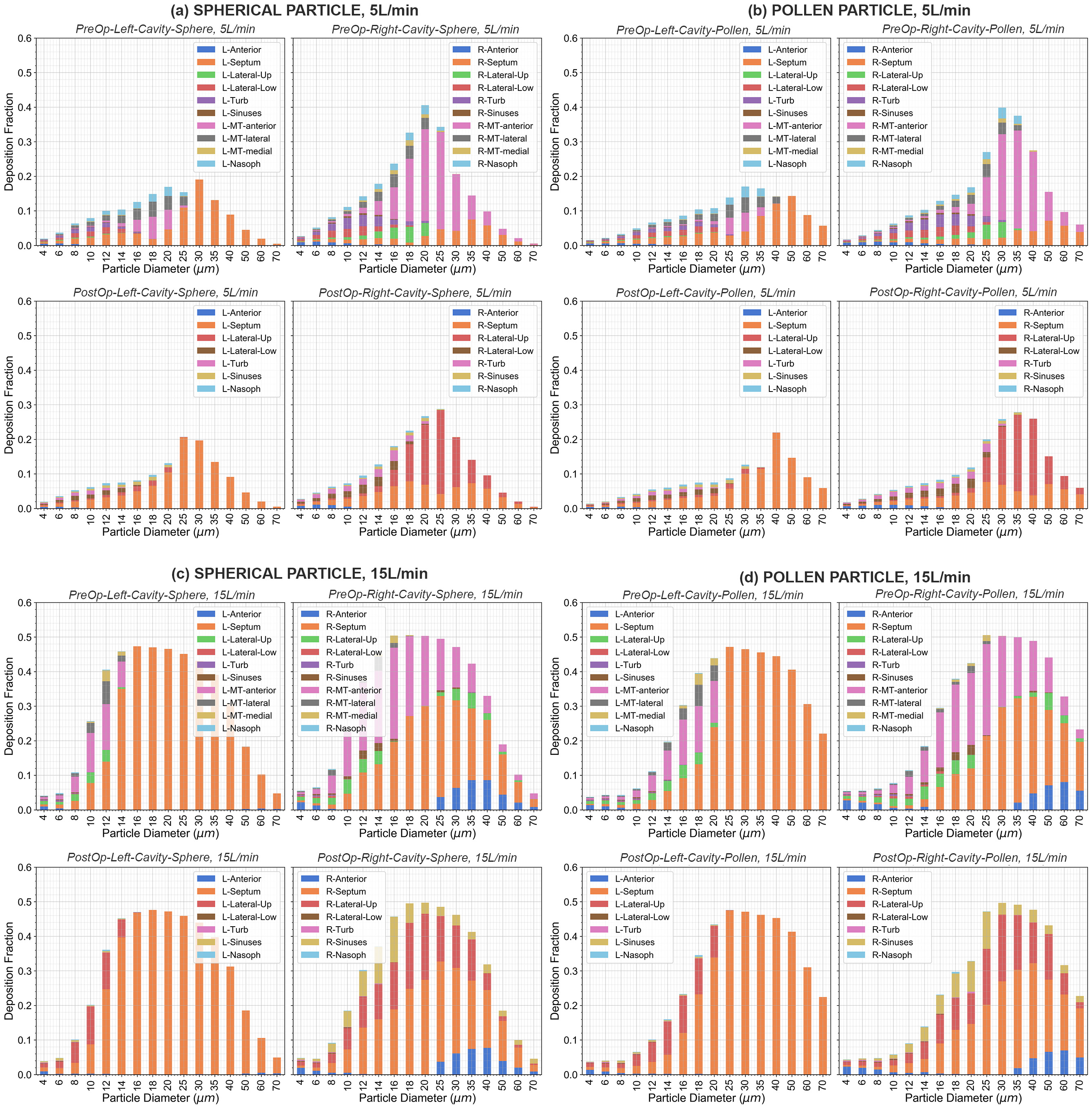}
	\caption{Deposition fraction, defined as a ratio of the number of particles depositing on a surface to the number introduced from the breathing region dome. Therefore, the surface deposition excludes particles escaping through the nasopharynx or not inhaled. (a) Spherical particles inhaled at 5L/min. (b) Pollen particles inhaled at 5L/min. (c) Spherical particles inhaled at 15L/min. (d) Pollen particles inhaled at 15L/min. Each subfigure depicts the deposition for PreOp (top panels), and PostOp (lower panels) models, separated in to left and right cavities.}
	\label{fig:depPlot}
\end{figure}

\clearpage
\newgeometry{margin=1.5cm}

\begin{figure}
	\centering
	\includegraphics[width=0.9\linewidth]{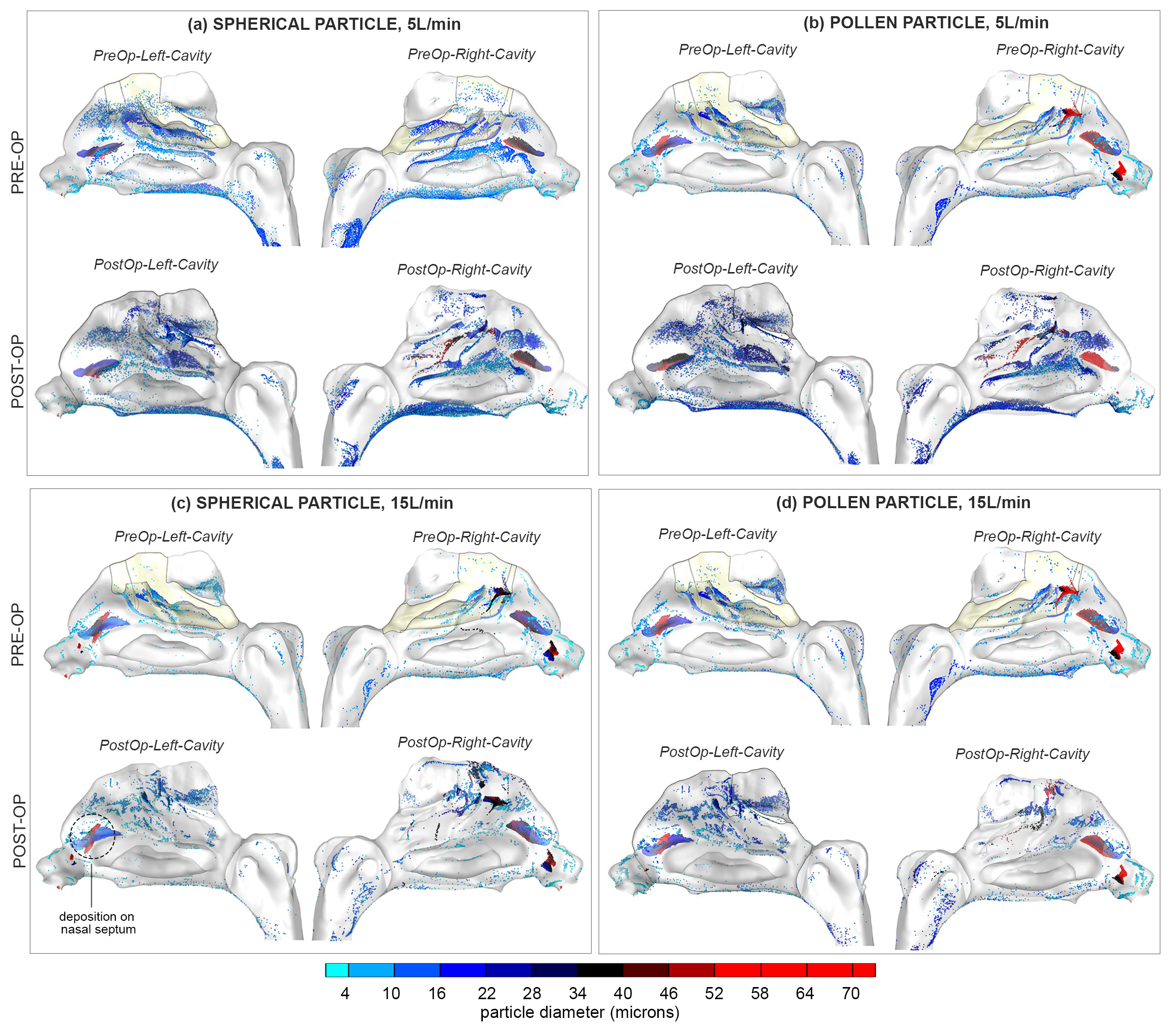}
	\caption{Deposition sites of inhaled particles. The paranasal sinuses have been excluded for improved visualisation of the main nasal passage deposition which was dominant. (a) Spherical particles inhaled at 5L/min. (b) Pollen particles inhaled at 5L/min. (c) Spherical particles inhaled at 15L/min. (d) Pollen particles inhaled at 15L/min. Each subfigure depicts the deposition for Pre-Op (upper locations), and Post-Op (lower locations) models, separated in to left and right cavities}.
	\label{fig:depSites}
\end{figure}

\clearpage
\begin{figure}
	\centering
	\includegraphics[width=0.9\linewidth]{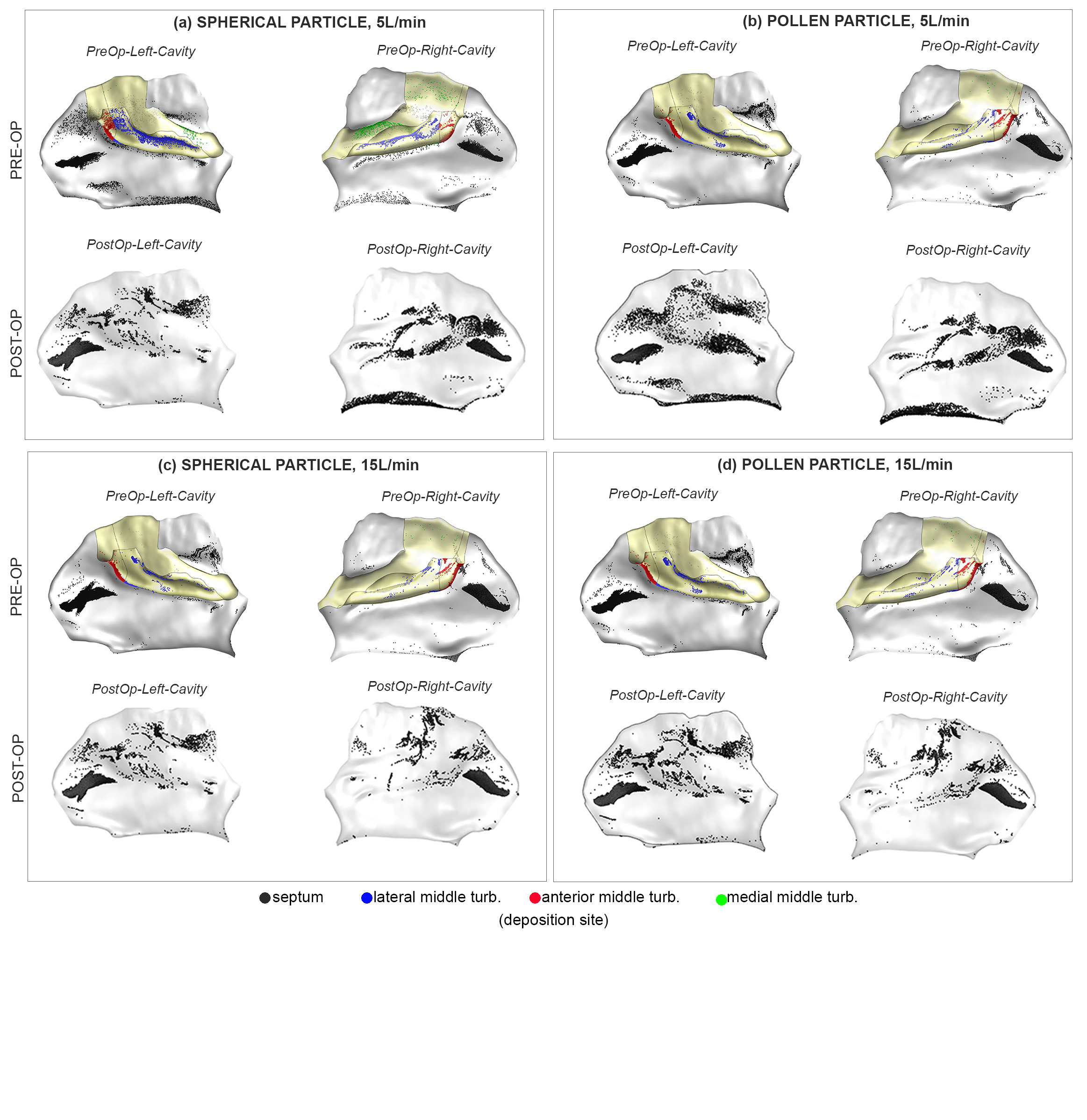}
	\caption{Deposition on the septum and middle turbinate (surface is coloured in yellow). Local deposition sites are coloured by surface: black - septum; blue - lateral middle turbinate; red - anterior middle turbinate; green - medial middle turbinate. (a) Spherical particles inhaled at 5L/min. (b) Pollen particles inhaled at 5L/min. (c) Spherical particles inhaled at 15L/min. (d) Pollen particles inhaled at 15L/min. Each subfigure depicts the deposition for Pre-Op (upper locations), and Post-Op (lower locations) models, separated in to left and right cavities.}
		\label{fig:depSeptum}
	\end{figure}
	
\clearpage
\begin{figure}
	\begin{subfigure}[b]{1\textwidth}
		\centering
		\includegraphics[width=0.7\linewidth]{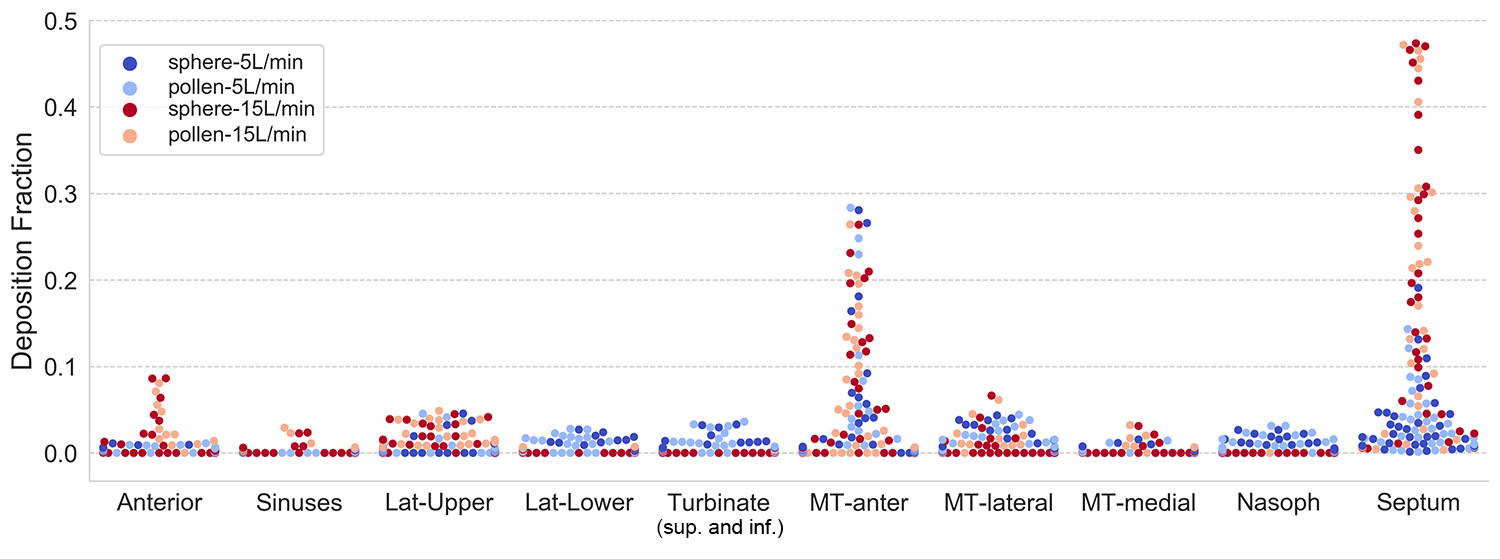}
		\caption{Deposition efficiency in Pre-Op model}
		\vspace*{2mm}
	\end{subfigure}
	~
	\begin{subfigure}[b]{1\textwidth}
		\centering
		\includegraphics[width=0.7\linewidth]{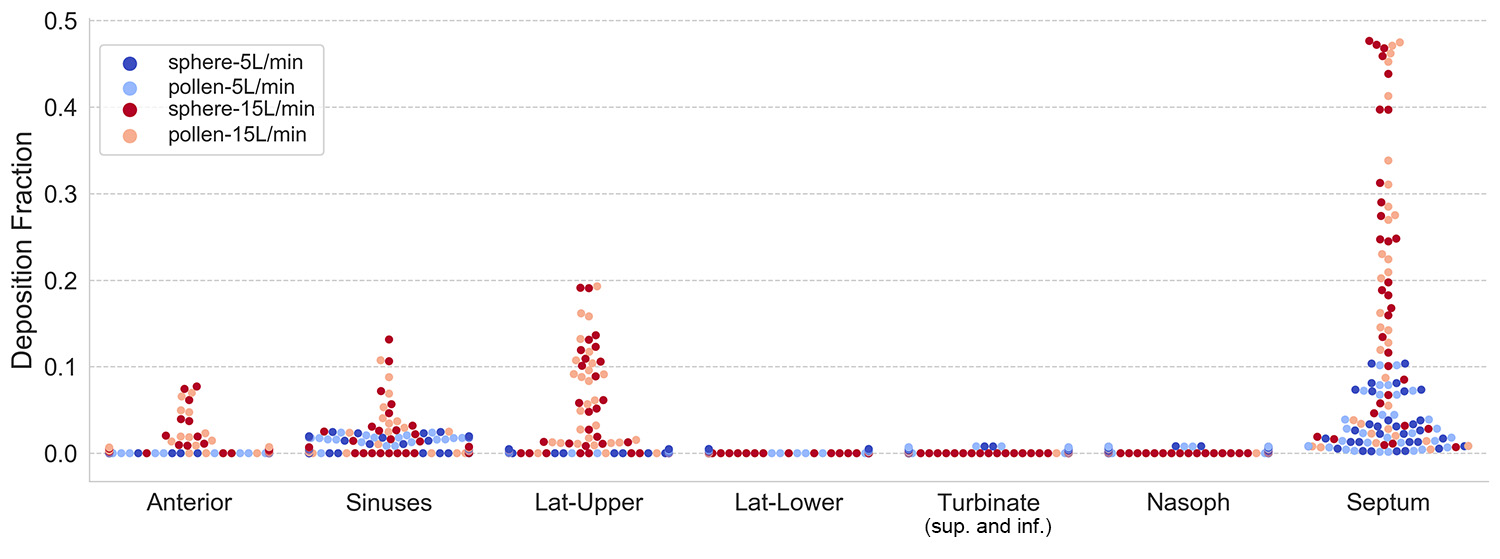}
		\caption{Deposition efficiency in Post-Op model}
		\vspace*{2mm}
	\end{subfigure}
	~
	\centering
	\begin{subfigure}[b]{1\textwidth}
		\centering
		\includegraphics[width=0.7\linewidth]{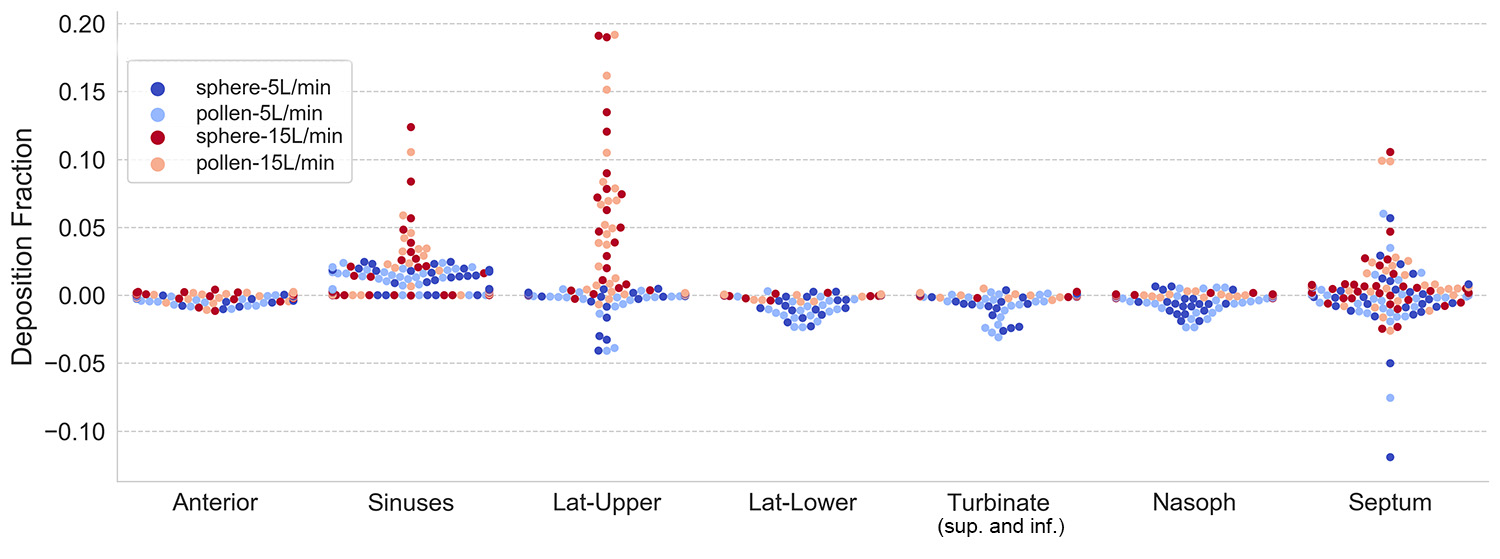}
		\caption{Change in deposition (Post-Op minus Pre-Op) coloured by particle type}
		\vspace*{2mm}
	\end{subfigure}
	
	~
	\centering
	\begin{subfigure}[b]{1\textwidth}
		\centering
		\includegraphics[width=0.7\linewidth]{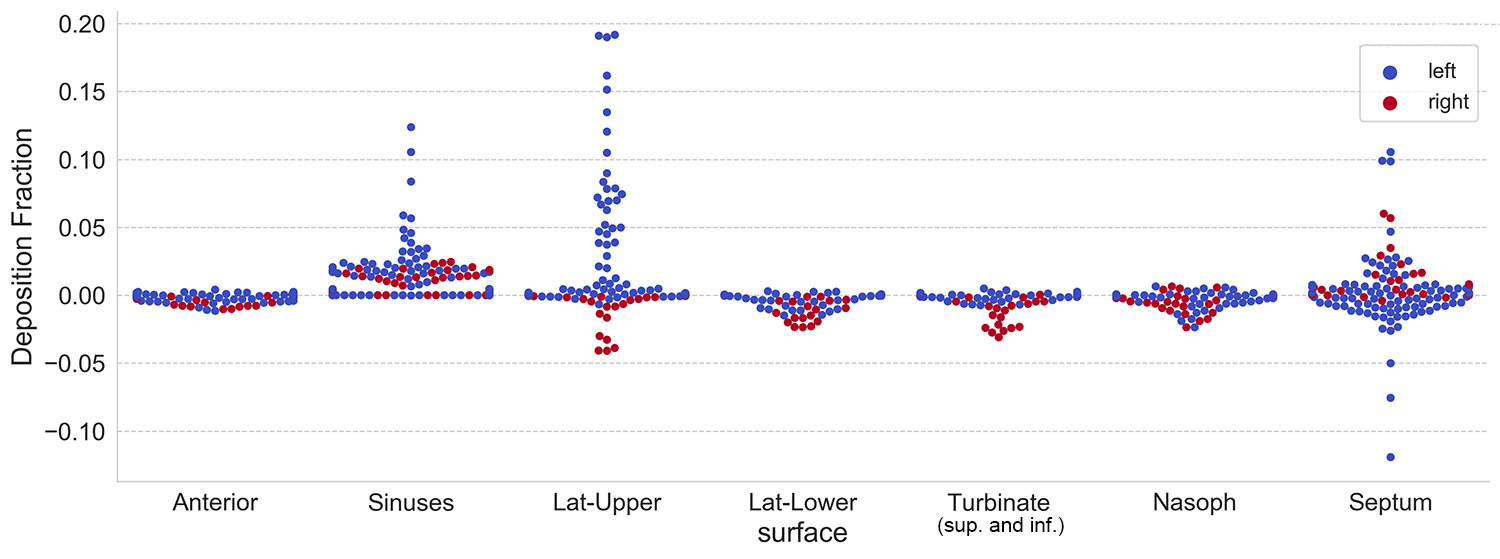}
		\caption{Change in deposition (Post-Op minus Pre-Op) coloured by nasal cavity side}
		\vspace*{2mm}
	\end{subfigure}

	\caption{Swarm plots for total deposition fractions in one cavity. Points on the plot having the same values of deposition will spread laterally.The height correlates with the deposition fraction, thus a higher point represents an occurrence of greater deposition fraction.} 
	\label{fig:swarm}
\end{figure}

\clearpage
\section*{Supplementary Material 1: Mesh Independence Analysis}
\begin{figure}[h!]
	\centering
	\includegraphics[width=0.6\linewidth]{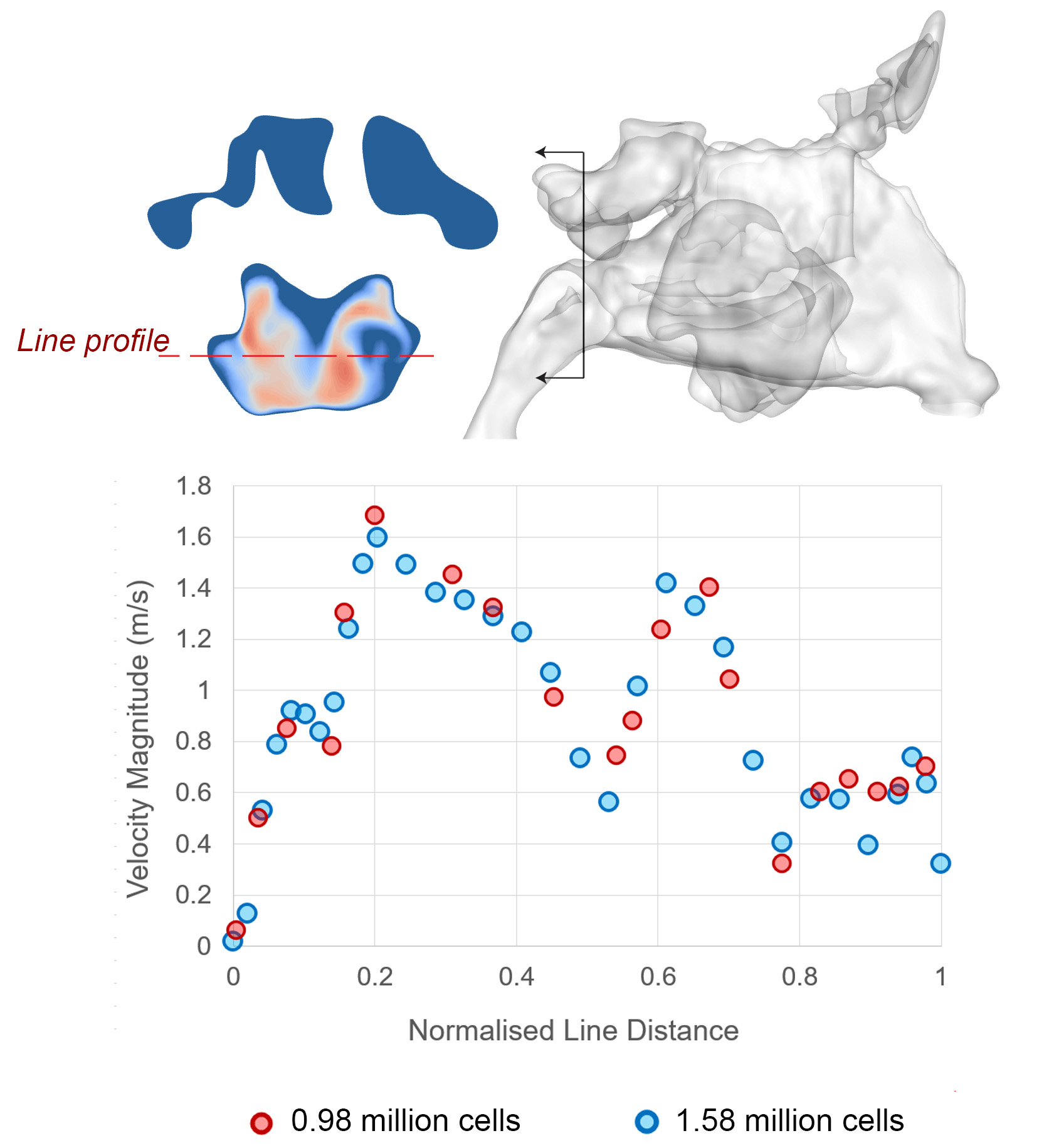}
	\caption*{Mesh independence analysis demonstrating the variation in velocity magnitude for a coarse mesh of 0.98 million polyhex-core cells, and a more refined mesh of 1.58 million polyhex-core cells.}
	\label{fig:meshInd}
\end{figure}

\end{document}